\documentclass[conference]{IEEEtran}
\usepackage[T1]{fontenc}
\IEEEoverridecommandlockouts
\usepackage{cite}
\usepackage{amsmath,amssymb,amsfonts}
\usepackage{algorithmic}
\usepackage{graphicx}
\usepackage{cite}
\usepackage{hyperref}
\usepackage{textcomp}
\usepackage{xcolor}
\usepackage{url} 
\def\BibTeX{{\rm B\kern-.05em{\sc i\kern-.025em b}\kern-.08em
    T\kern-.1667em\lower.7ex\hbox{E}\kern-.125emX}}
\begin{document}
% --- Preprint notice page ---
\begin{titlepage}
    \centering
    \vspace*{\fill}
    {\Large © 2025 IEEE.
Personal use of this material is permitted.  Permission from IEEE must be obtained for all other uses, in any current or future media, including reprinting/republishing this material for advertising or promotional purposes, creating new collective works, for resale or redistribution to servers or lists, or reuse of any copyrighted component of this work in other works.

 This is the author’s version of the article that has been accepted for publication in IEEE ISSE 2025. 
The final version is available at IEEE Xplore.

DOI: https://doi.org/10.1109/ISSE65546.2025.11369983

}
    \vspace*{\fill}
\end{titlepage}
\title{LLM-Assisted Semantic Alignment and Integration in Collaborative Model-Based Systems Engineering Using SysML v2\\
}

\author{
\IEEEauthorblockN{1\textsuperscript{st} Zirui Li}
\IEEEauthorblockA{\textit{Product and Systems Engineering Group} \\
\textit{Technische Universität Ilmenau}\\
Ilmenau, Germany \\
https://orcid.org/0009-0007-7983-7901}

\and
\IEEEauthorblockN{2\textsuperscript{nd} Stephan Husung}
\IEEEauthorblockA{\textit{Product and Systems Engineering Group} \\
\textit{Technische Universität Ilmenau}\\
Ilmenau, Germany \\
https://orcid.org/0000-0003-0131-5664}

\and
\IEEEauthorblockN{3\textsuperscript{rd} Haoze Wang}
\IEEEauthorblockA{\textit{Product and Systems Engineering Group} \\
\textit{Technische Universität Ilmenau}\\
Ilmenau, Germany \\
https://orcid.org/0009-0009-9826-9027}
}
\maketitle

\begin{abstract}
Cross-organizational collaboration in Model-Based Systems Engineering (MBSE) faces many challenges in achieving semantic alignment across independently developed system models. SysML v2 introduces enhanced structural modularity and formal semantics, offering a stronger foundation for interoperable modeling. Meanwhile, GPT-based Large Language Models (LLMs) provide new capabilities for assisting model understanding and integration. This paper proposes a structured, prompt-driven approach for LLM-assisted semantic alignment of SysML v2 models. The core contribution lies in the iterative development of an alignment approach and interaction prompts, incorporating model extraction, semantic matching, and verification. The approach leverages SysML v2 constructs such as alias, import, and metadata extensions to support traceable, soft alignment integration. It is demonstrated with a GPT-based LLM through an example of a measurement system. Benefits and limitations are discussed.
\end{abstract}


\section{Introduction}
In the development of modern complex systems, Model-Based Systems Engineering (MBSE) approach has become increasingly important \cite{estefan2007survey}. It emphasizes the use of formal or semi-formal modeling throughout the system lifecycle to model elements of systems, analyze them, and use them for different tasks \cite{Hick.2019}. This approach not only improves the systematic engineering process but also provides a structured manner \cite{Boelsen.2025,Zhou.2025}, which has the potential to simplify cross-company collaboration. 

In a typical collaboration scenario where Original Equipment Manufacturer (OEM) and suppliers develop systems collaboratively, subsystem models are often created independently by different partners and then integrated by OEM to enable holistic analysis and assessment \cite{Li.2024,Lu.2007,Duehr.2019}. Different methods can be used for modeling (see e.g. \cite{Morkevicius.2023,Bohm.2021}). The differences in modeling methods and understanding of the engineers exacerbate inconsistencies in model structure and semantics, particularly in the early stages of system development when system maturity is low and design iterations are frequent \cite{ElHamlaoui.2021,Feldmann.2019,estefan2007survey}. Aligning the intended subsystems requirements (Top-down) with as-is specifications of subsystems (Bottom-up) remains highly dependent on manual processing\cite{Li.2023}. SysML v2, as a new generation of system modeling language, provides formalized semantics, stronger text modeling support, and structural extension capabilities \cite{Boelsen.2025,Bajaj.2022} and thus provides a good basis for improving collaborative MBSE.

Recently, with the development of Large Language Models (LLMs) such as ChatGPT, Artificial Intelligence (AI) has gradually been applied to the field of engineering modeling to support e.g. understanding, transformation, and generation of model information \cite{Ghanawi.2024,Zhang.2024,Kulkarni.2024}. Nevertheless, LLM still faces challenges such as insufficient understanding, uncertainty in output, and a lack of verification and tracking mechanisms for the engineering modeling \cite{Zhang.2024,Hadi.2023,Hollender.2024,Clariso.2023,White.2024}.

Therefore, the core research question addressed in this paper is: 
\begin{itemize}
\item How can LLM technology and SysML v2 constructs be combined to enable semantic alignment and integration of cross-organizational MBSE models? 
\item How can a structured alignment approach be developed to support company-specific semantic extensions, enabling LLM to perform more context-aware model understanding and integration?
\end{itemize}

This paper bases on previous research work \cite{StephanHusungZiruiLiFaizanFaheem.2024} and focuses on exploring a LLM-assisted approach related to the research question and the introduction of modeling extensions to achieve model alignment and integration.

\section{State of The Art}

\subsection{Collaboration in MBSE}
Towards 2035, INCOSE states that the future of Systems Engineering (SE) will increasingly rely on model-driven ways of working to meet the development requirements of complex systems across domains and organizations\cite{SystemsEngineeringVision2035.6172025}. In this context, SE focuses on issues such as modeling languages semantic interoperability, tool chains and standardization of methodologies\cite{Miller.2022}. 

In current collaborative development practices, there are structural heterogeneity and semantic misalignment among models because different teams usually use different modeling method (see e.g. \cite{Morkevicius.2023,Bohm.2021}), modeling tools, and domain abstraction approaches. This problem is especially apparent in cross-organizational scenarios such as OEM-supplier co-design. Studies have shown that even the same artifacts for the same product may be given different structural and semantic representations in the modeling process of different companies\cite{Hamlaoui.2018}.

Consistency management is another essential aspect of collaborative modeling. Feldmann et al. \cite{Feldmann.2019} proposed a dedicated graphical modeling language to define dependencies and consistency rules between models, enabling automated checks throughout the model lifecycle. Lu et al. \cite{lu2023detection} transformed SysML models into OWL ontologies and used semantic reasoning to detect logical conflicts, achieving semantic-level consistency checking. Harder et al. \cite{harder2025interpreting} combined SysML diagrams with OWL reasoning to create a unified modeling and inference framework, broadening the use of ontology reasoning in engineering modeling. However, these solutions typically require extensive upfront modeling discipline and lack flexibility when scaling across heterogeneous domains and companies.

\subsection{Application of SysML v2}
With the growing complexity of SE, traditional SysML v1 shows many limitations: limited expressiveness, challenges in automating graphical models, and semantic ambiguities resulting from implementation as a UML profile. These challenges have driven the development and optimization of a new generation of modeling languages to meet the demands in MBSE \cite{Gray.2021}. The introduction of SysML v2 aims to enhance its capabilities in terms of modeling expressiveness, tool interoperability, and automated verification through semantic restructuring and language architecture, features such as text-based modeling, modular structure, and a unified semantic core to enhance its overall capabilities in terms of expressiveness, tool interoperability, and model validation support \cite{Gray.2021,Jansen.2022}.

Previous studies have investigated SysML v2 from multiple perspectives. Meanwhile, the formal semantics emphasized by SysML v2 support integration with verification tools, enabling model translation for contract checking and reachability analysis using frameworks like Gamma and Imandra \cite{Molnar.2024}. From a language engineering view, research has addressed syntax consistency and maintainability, noting that SysML v2 still requires improvements in grammatical clarity and semantic documentation \cite{Jansen.2022}. Furthermore, research targeting industrial scenarios includes proposed standardization guidelines for modeling in mechanical systems \cite{Boelsen.2025} and the exploration of new modeling meta-models for product line variability management \cite{Epp.2023}. Friedenthal et al. \cite{Friedenthal.2023} summarized the potential of SysML v2 to support the ‘single source of truth’ practice from the perspective of MBSE system evolution, and pointed out that lifecycle modeling, semantic consistency control, and model reuse will be key directions for future research.

\subsection{LLM-assisted Engineering}
In recent years, research into the application of LLMs in the field of MBSE has gradually emerged, demonstrating significant potential in areas such as model generation, semantic understanding, and system integration. The INCOSE-published ‘Systems Engineering Vision 2035’ emphasizes that future SE environments will be more uncertain, dynamic, and interdisciplinary, and that SE can leverage technologies such as AI to enhance its adaptability and sustainability \cite{SystemsEngineeringVision2035.6172025,Miller.2022, Lipsinic.2025}. This paper reviews the current state of art and future directions for LLMs in SE from three perspectives: prompt engineering, AI and domain knowledge integration, and challenges in the engineering context.

Recent studies have investigated how LLMs can be guided through prompt engineering to perform structured engineering tasks. Researchers have found that, with well-designed prompts, LLMs are capable of generating system architecture elements without requiring specialized domain fine-tuning \cite{Kulkarni.2024}. Instead of relying on unstructured dialogue, model-driven prompt engineering introduces formalization through domain-specific languages (DSLs), allowing for greater consistency, reuse, and version control across tools and environments \cite{Clariso.2023}. Complementary efforts, such as the development of prompt pattern catalogs, provide reusable templates for interacting with LLMs in tasks like requirements classification and architecture design in the early phase \cite{White.2024}.

To improve semantic precision and contextual alignment, recent studies have investigated integrating LLMs with structured engineering knowledge bases. Gauthier et al. \cite{mbse-ai.integration25} proposed a hybrid approach that combined engineering ontologies and Retrieval Augmented Generation (RAG) to produce domain-specific content. In other practical applications, LLMs have been tested for their ability to extract modeling elements directly from normative standards, reducing the manual modeling effort  and supporting the generation of SysML models for integration into MBSE\cite{Ghanawi.2024}.

Despite these advances, several limitations hinder the practical adoption of LLMs in engineering contexts. First, the lack of domain-specific training data and limited contextual understanding make it difficult for LLMs to interpret domain-specific semantics, especially in SE modeling scenarios \cite{Topcu.2025, Kulkarni.2024,Hollender.2024b}. Second, LLMs outputs can be unreliable, with risks of hallucination or semantic inconsistency, which compromise their utility in the engineering process\cite{White.2024, Ghanawi.2024,Hollender.2024b}. Third, current implementations typically lack mechanisms for traceability, validation, and version control, making them difficult to integrate into rigorous engineering processes \cite{Clariso.2023, SystemsEngineeringVision2035.6172025}.

These limitations highlight the demand for more structured and verifiable integration approaches that align LLM capabilities with the demands of MBSE.

\section{Scientific Approach to LLM-Assisted Model Alignment}
This paper proposes a systematic approach that combines LLM and SysML v2 constructs to assist in model alignment and integration in collaborative MBSE. The proposed approach is organized into multiple structured stages, with alternating AI support and human verification. It emphasizes process orientation, semantic control, and structured prompt interaction to support traceability and model consistency.

Compared to traditional manual alignment process, this approach aims not only to improve efficiency but also to enhance the semantic accuracy, structural stability, and traceability of aligned model outputs from LLM.

\subsection{Challenges in LLM-Assisted Model Alignment}\label{AA}

Before developing an AI-assisted model alignment approach, it is necessary first to clarify the requirements that the approach must meet. These requirements not only determine the verification criteria for approach design but also reflect the current engineering practices and technical challenges faced in LLM applications.
\begin{itemize}
\item	\textbf{Efficient model alignment capability}: Existing manual or rule-based alignment approaches often encounter issues such as model complexity and semantic ambiguity \cite{Feldmann.2019}. LLM can assist in extracting semantically similar items from natural language documents and part of existing models \cite{Ghanawi.2024,Zhang.2024}, but should address challenges such as semantic ambiguity and the difficulty of mapping on model hierarchy level.
\item	\textbf{Output validity, repeatability, and semantic consistency}: LLM often produces non-deterministic results (e.g., different outcomes from the same prompt run multiple times) and may generate hallucinated or inconsistent outputs \cite{Hadi.2023,Hollender.2024,Clariso.2023}. Consequently, output verification procedures and structural control mechanisms are required to ensure that the alignment process does not alter or compromise the original model’s structural integrity and semantic correctness.
\item	\textbf{Output traceability}: In engineering environments, all outputs must be traceable and verifiable. LLM outputs are often black boxes \cite{Kulkarni.2024,White.2024}, so mechanisms must be designed to support input-output mapping records, confidence annotations, and user interaction traces to enable human verification and subsequent maintenance.
\end{itemize}
In response to these requirements and challenges, a prompt-driven approach was proposed to bridge the gap between the limitations of LLMs and the demands of collaborative MBSE.

\subsection{Model Integration Concepts}\label{sec:alignment}
To formulate effective LLM-supported approaches, it is necessary to briefly revisit possible integration concepts discussed in previous work \cite{StephanHusungZiruiLiFaizanFaheem.2024} —"unified modeling", "transformation-based integration", and "soft alignment" -- with particular attention to their suitability for LLM application.

"Unified modeling" aims to define a unified methodology and model structure during collaboration. While beneficial in theory, this is rarely feasible in practice due to organizational company-specific modeling requirements. Additionally, research on the semantic layer of unified models (such as ontology-driven approaches like CASCaDE \cite{CASCaDEHomepage.632025}) is still in its early stages and requires further investigation, limiting LLMs' potential contributions in this context.

"Transformation-based integration" focuses on converting models across modeling methodologies (e.g., SPES \cite{Bohm.2021} to MagicGrid \cite{Morkevicius.2023}). Although partially automatable, the absence of standardized SysML v2 methodologies and models restricts LLM learning and execution. Current tools and methods focus on syntax-level migration from SysML v1 to v2 \cite{Friedenthal.2023, Zhou.2024}, while semantic alignment remains underdeveloped.

"Soft alignment" creates new alignment packages that map elements between OEM and supplier models, while retaining the original structure of both. Instead of enforcing a unified structure, this approach supports independent development by mapping relevant elements through lightweight extension libraries of existing SysML constructs (e.g. allocations). Although it requires initial manual configuration of semantic libraries, it leverages LLM strengths in natural language processing and semantic reasoning. LLMs can support traceable, collaborative alignment through suggestion and refinement.

In summary, "soft alignment" currently offers the most feasible and LLM-compatible concept. Its focus on semantic mapping rather than structural modification, which aligns well with iterative, cross-organizational integration scenarios.

\subsection{Iterative Refinement of the Integration Approach}\label{sec:Refinement}
Developing an LLM-assisted approach for semantic alignment and integration of SysML v2 models faces challenges. The non-deterministic nature of LLM makes their behavior in engineering contexts inherently difficult to predict and control \cite{Kulkarni.2024,Hadi.2023,Clariso.2023}. Furthermore, the complex and evolving semantics of collaborative MBSE models require a process that is not only accurate but also traceable, verifiable, and acceptable within the engineering process. 

To address these challenges, this work adopts an iterative design approach based on agile test process development \cite{Baumgartner.2023} and Design Research Methodology (DRM) \cite{Blessing.2009}. The process was developed and refined using an OEM and supplier SysML v2 model example, supported by a domain-specific extension library for soft alignment. Structured prompts and staged user interactions were used to coordinate the ChatGPT-based assistant (GPT-4o) during the alignment process.

Initial experiments with direct LLM prompting quickly exposed critical limitations, including inconsistent outputs, lack of traceability, and limited user control. This motivated the iterative refinement toward a structured, stage-based process that introduced clear phases for syntax verification, model summarization, match generation, and result export.

In line with agile testing principles, each iteration incorporated early and continuous verification activities. Intermediate results were manually reviewed and assessed for semantic accuracy, structural consistency, and alignment confidence. Issues observed in the output were systematically analyzed, and corresponding prompt modifications were made.

Through multiple refinement cycles, the process was enhanced to improve semantic depth, consistency, and transparency. Completeness checks were added to the model extraction phase; unmatched elements were explicitly reported in the matching phase; and verifications included rationale, confidence scores, and compliance with SysML semantics. Additive model packages were generated using proper alias/import structures to preserve the original model structure, and results were exported in both machine-readable (JSON) and human-readable formats. Structured annotations were also embedded to convey alignment rationale.

The result of these iterations showed improved stability, traceability, and output consistency across repeated runs. However, the test result was limited to a small set of models and semantics. Broader verification remains necessary to confirm the approach's ability. Building on these insights, a structured LLM-assisted model integration approach is summarized in the following section.

\section{Result}

\subsection{LLM-assisted Integration Approach}
Building upon the insights of the iterative refinement of the integration approach (see subsection \ref{sec:Refinement}), the following section summarizes the staged process for the alignment and key mechanisms designed to support reliable and traceable semantic alignment of SysML v2 models. 

The approach builds on the concept of soft alignment, which avoids unified modeling and instead promotes a loosely mapped integration mechanism (see subsection \ref{sec:Refinement}). This approach is extended through the structural and semantic constructs of SysML v2 and the interactive potential of LLMs. Specifically, the process combines:
\begin{itemize}
    \item Structural reference and element reuse via SysML v2 constructs such as alias, public/private import, and package structures;
    \item Semantic extensibility through a user-defined extension library for identifying and differentiating alignment options;
    \item Prompt-driven generation by structuring the LLM interaction into staged tasks with user checkpoints and verification steps.
\end{itemize}	

Based on the investigation, the following integration approach is proposed:
\begin{itemize}

\item \textbf{Additive Modeling}: Rather than directly modifying the original model content, LLM generates extended content in an additionally created package and references the original model elements via the SysML v2 'import' mechanism. This approach prevents the destruction of the model structure and supports multiple rounds of incremental alignment.
\item \textbf{Staged Process}: The process involves structured sequences and incorporates LLM interactive guidance, human verification and intervention at each stage to ensure the accuracy and engineering acceptability of the output.
\item \textbf{Confidence-Scored Mapping Suggestions}: Each model mapping recommendation is accompanied by an LLM-assigned confidence score to aid engineers in interpretation;
\item \textbf{Mapping Verification}: All suggested mappings undergo a secondary verification based on the SysML syntax and semantic documentation;
\item \textbf{Coverage Check}: LLM must perform and report a coverage check to confirm that all model elements and prior stage outputs have been fully and correctly processed, ensuring traceability and preventing silent omissions.
\item \textbf{Standardized Output Format}: It is particularly suitable for the semantic pre-processing stage before model understanding. If formats are not standardized, LLM may produce unstable output structures and semantic confusion during model element extraction and semantic classification, increasing the cost of manual verification. Adopting a unified structure such as JSON or specific syntax markup ensures that LLM outputs have structural consistency, facilitating subsequent processing and verification.
\item \textbf{Structured Comment Support}: The generated SysML v2 model includes structured comment elements attached to relevant model items, providing the rationale and intended meaning of alignment decisions. These comment elements are designed to enhance user understanding of the alignment context directly within the model.

\end{itemize}

The approach described above can be realized with a system prompt, inspired by the structured prompt pattern proposed in \cite{White.2024}. The prompt emphasizes a staged process to drive model semantic understanding and structural generation, while incorporating user interaction confirmation throughout.

\subsection{Process Overview}
\label{sec:ProcessOverview}

\begin{figure*}[t]
\centering
\includegraphics[width=\textwidth, keepaspectratio]{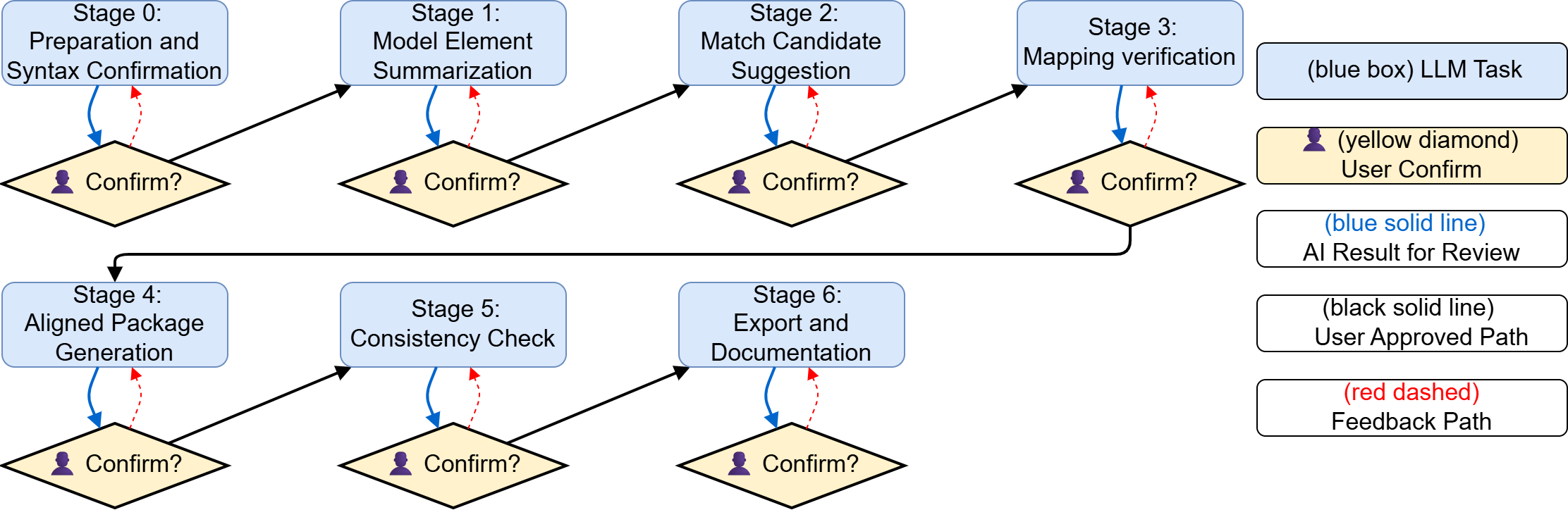}
\caption{LLM-assisted SysML v2 Model Alignment Process}
\label{fig:alignment}
\end{figure*}

Figure ~\ref{fig:alignment} illustrates the LLM-assisted SysML v2 model alignment process, emphasizing structural stability and semantic clarity. The process is divided into seven stages. Each stage requires explicit user confirmation to proceed; otherwise, the system will iteratively optimize within the current stage. The stages are as follows:
\begin{itemize}
\item \textbf{Stage 0 - Preparation and Syntax Confirmation}: Users must provide textual model description (.txt), optional Unique Identifier (UID) for model information, and optional semantic extension content (SysML v2 extension library). The LLM verifies the format, parses the contextual structure, and generates preliminary analysis feedback. Additionally, to validate whether the LLM can correctly identify and utilize the user-uploaded semantic extensions, the LLM should automatically generate a model alignment example based on the extension library;
\item \textbf{Stage 1 - Model Element Summarization}: The LLM extracts structural definitions, usage, interfaces, requirements, and semantic annotations, and outputs them uniformly in JSON format or an intermediate representation structure;
\item \textbf{Stage 2 - Match Candidate Suggestion}: Based on information such as naming similarity, interface matching, extended semantics, or contextual tags, candidate matching pairs are proposed. If the user intends to apply specific SysML v2 constructs (e.g., using allocation to assign subsystem behavior to an overall system use case), this intent should be explicitly specified in the prompt. Without such guidance, the LLM may generate results that lack semantic focus, thereby reducing consistency and controllability of the outputs.
\item \textbf{Stage 3 - Mapping Verification}: The LLM performs semantic consistency analysis on the matching pairs, identifies structural conflicts or inconsistencies in abstraction levels, followed by user verification of their semantic appropriateness;
\item \textbf{Stage 4 - Aligned Package Generation}: Create an alignment result package using references (e.g., private/public import), semantic relationships (e.g., specialize, connection, allocation) and extension library to maintain the independence of the original model structure;
\item \textbf{Stage 5 - Consistency Check}: Verify model structure, reference scope, semantic relationships, and extended semantic consistency;
\item \textbf{Stage 6 - Export and Documentation}: Export the integrated model, matching logs, and a list of potential issue diagnoses.
\end{itemize}

This process is designed to support semantic alignment and abstract mapping in SysML v2 through semantics such as specialization, subset, redefine, extended metadata, semantic annotations, and allocation. These semantics suggest potential for scalability, but further investigations are needed to confirm its effectiveness in complex or large-scale industrial integration scenarios.

\subsection{Semantic alignment instance}
To achieve semantic alignment between independently developed SysML v2 models, this subsection illustrate an instance of soft alignment concepts that preserves the structural independence of original models. Rather than enforcing structural unification, the alignment leverages existing SysML v2 constructs—such as alias and import, combined with metadata-based extensions to enhance interpretability and traceability. These constructs allow LLMs to suggest semantic alignments without modifying original model structures.

\begin{figure*}[t]
\centering
\includegraphics[width=\textwidth, keepaspectratio]{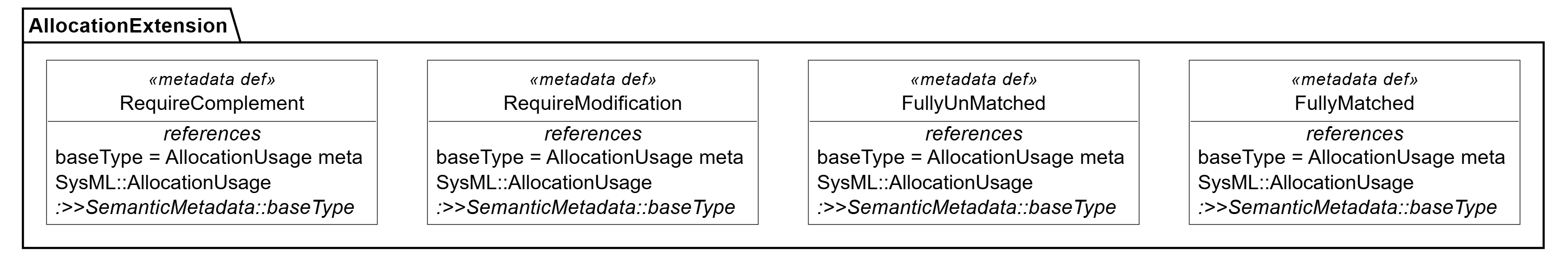}
\caption{Excerpt of Alignment Extension Library}
\label{fig:extension}
\end{figure*}

In this context, alias and import are used to support lightweight referencing and reuse of elements between models. In SysML v2, the alias construct provides a lightweight name binding mechanism, allowing semantic equivalence mappings between model elements without modifying the original structure \cite{.642025}. This supports naming consistency, offering a feasible approach for additive model alignment in cross-companies modeling. The SysML import construct enables visibility of elements  via public or private import, enabling external reuse while preserving encapsulation, which is useful for maintaining structural independence in collaborative modeling and alignment. However, as SysML v2 and its supporting tools are still evolving, the practical implementation and effectiveness of these constructs require further evaluation and research \cite{Epp.2023,Gray.2021}. To further support alignment, a semantic extension library based on the SemanticMetadata package was developed. It extends AllocationUsage with metadata to label match result—such as ‘FullyMatched’, ‘RequireComplement’, ‘RequireModification’, and ‘FullyUnmatched’, as shown in Figure~\ref{fig:extension}, thereby improving the interpretability and traceability of LLM-generated alignment outputs.

\subsection{Prompt-driven Realization and Verification}

To implement the process execution logic, a systematic template of the prompt was designed based on alias and extension alignment, which includes the following points:
\begin{itemize}
    \item only alias and extension are used for model alignment, without rewriting the original model structure,
    \item all alignment suggestions are provided in JSON format, including confidence score and explanation,
    \item during the stages (see subsection \ref{sec:ProcessOverview}), the LLM should perform self-checks for structural and semantic consistency, and validate format and references using SysML v2 syntax rules and
    \item user confirmation points are set at each phase to control process transparency and accuracy.
\end{itemize}
In Stage 0 \textit{Preparation and Syntax Confirmation}, the LLM initially misinterpreted the user-provided semantic extension library. It incorrectly introduced \textit{\#FullyMatched} with semantic wrong structures in stage summary outputs. This behavior indicated that the LLM had misunderstood the correct application format of the extension.

To address this, the user intervened and provided the following prompt to clarify the semantic rule:

\begin{quote}
\textit{allocation extension is wrong. right form: \#FullyMatched allocation element1 to element2. element cannot be definitions.}
\end{quote}

After receiving this clarification, the LLM corrected its behavior and consistently applied the correct usage pattern in subsequent stages without requiring further instructions. Its outputs became stable, context-aware, and syntactically valid, with user interaction limited to confirmation.

\begin{figure*}[t]
\centering
\includegraphics[width=\textwidth, keepaspectratio]{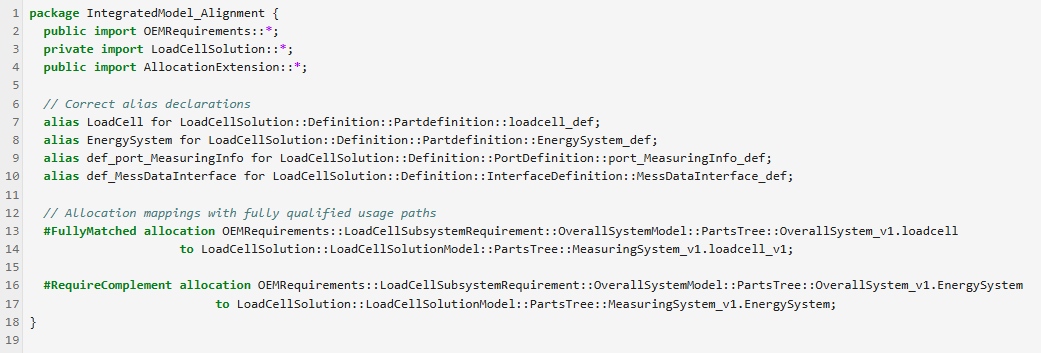}
\caption{Example Model Alignment Results}
\label{fig:example}
\end{figure*}

This experiment illustrates how minimal human intervention can rectify early-stage semantic deviations in a structured prompt-driven workflow.
Figure~\ref{fig:example} illustrates alignment results, which demonstrate process completeness, output traceability, and alignment transparency across test scenarios derived from OEM and Supplier SysML v2 example model in \texttt{.txt} file. These results highlight the LLM's ability to generalize from minimal human correction, supporting effective alignment under structured prompting. Limitations and observations from these evaluations are discussed in section ~\ref{dis} to inform future refinements and broader validation efforts.

The complete prompt structure is available in the GitHub project repository under the file \path{prompts/sysmlv2_alignment_process.md}~\cite{GitHub.6202025}.

To illustrate the practical application of the proposed approach, example model alignment results are provided in Figure ~\ref{fig:example} and GitHub \path{prompts/examples/results/IntegratedModel_Alignment.txt}~\cite{GitHub.6202025}. The verification focuses on demonstrating process completeness, output traceability, and alignment transparency across test scenarios derived from OEM and supplier SysML v2 models. 

\section{Discussion}\label{dis}
Although the prompt-driven model alignment process proposed in this contribution demonstrates certain advantages in terms of structural and semantic transparency, there are still limitations in deep integration, interoperability, and maintainability.

Firstly, the current method primarily focuses on identifying structural consistency at the naming level, effectively handling structural alignment and alias mapping. However, it has not yet fully addressed deeper semantic discrepancies such as inconsistencies at the abstraction level, differences in interface granularity, or mismatches in design intent. Therefore, future research should introduce alignment approaches targeting interface structure and behavioral semantics to enhance adaptability to complex heterogeneous models.

Second, the system prompt structure currently requires manual configuration each time. How to construct stable, reusable prompt modules or persistent fine-tuned models to achieve stable output with prompts is a key direction for improving usage efficiency and consistency.
Moreover, LLMs’ understanding of complex semantic extensions remains unstable. During initial use, inconsistencies in applying custom extension libraries can negatively affect the reliability of alignment outputs. To address this, one potential enhancement is the incorporation of ontologies as structured semantic references. These ontological structures could complement prompt design or act as consistency constraints, helping LLMs generate more grounded and verifiable outputs\cite{Zhang.2024, mbse-ai.integration25, CASCaDEHomepage.632025}. How to introduce auxiliary prompt generation, semantic constraint prompts, or semi-automatic model integration while maintaining user control, will be one of the research directions for deepening semantic alignment and improving user operability.

Additionally, the current system lacks a transparent and systematic feedback mechanism for explaining why certain mappings are suggested or rejected. This limits user oversight and interpretability of the alignment process. Introducing structured reasoning elements or justification logs would improve user trust, controllability and traceability of model.

To extend applicability in industrial scenarios, improvements are needed in model version control, toolchain integration, and automation. In the future, it may be possible to further integrate official SysML v2 API to achieve automated processes. Additionally, REST API interfaces could be introduced to enable data interoperability between front-end and back-end systems, thereby supporting model version management, prompt module reuse, and state preservation within an enterprise-level tool chain.

In practice, balancing the level of prompt completeness with the limitations of LLM context handling and generalization capabilities remains an open challenge. Highly detailed prompts with extensive stage-specific rules can improve structural consistency, but can increase processing latency, cognitive overload on the LLM, and output instability due to attention degradation and token limits\cite{liu2023lost}. In contrast, too-simple prompts tend to result in hallucinations, incomplete outputs, and lower repeatability \cite{White.2024, Ghanawi.2024,Hollender.2024b}. Our current process applies structured prompts selectively, with high completeness for extraction, candidate matching, and model generation stages, while allowing more interactive flexibility in validation and reporting phases. Nevertheless, further research is needed to systematically optimize prompt engineering for industrial-grade robustness.

Overall, the method demonstrates the potential of combining prompt-driven control, semantic extensions, and human-in-the-loop interaction for SysML v2 model integration, but still requires advancements in deep semantic alignment, version control, and API-level integration.

\section{Conclusion and Future Work}

This paper focuses on the task of  integration of SysML v2 models with LLM assistance, proposing a prompt-driven semantic alignment system method for cross-company collaborative MBSE. By combining the use of alias, import, and extended semantic library, the method supports incremental modeling and semantic alignment between models without compromising the original model structure.

The core advantages of this approach are:
\begin{itemize}
    \item By employing a clear system prompt template design, the alignment process is divided into distinct phases, enhancing user control and process transparency;
\item The introduction of a semantic extension library provides a relative general semantic mapping intermediary for multi-companies’ collaboration scenarios, enabling collaborative partners to share domain knowledge and build common semantics without unifying modeling methodologies, while also providing a sustainably extendable semantic basis for the model alignment process.
\item Support for multi-round matching suggestion generation and feedback iteration results in a stable, structure-independent alignment package structure.
\end{itemize}

Preliminary tests in a typical measurement system collaboration scenario indicates potential feasibility in terms of model structural consistency, semantic clarity, and user interaction efficiency. However, the current approach does not yet cover all semantic-level alignment requirements, the prompt mechanism still relies on the user's experience and guidance in prompt construction, and the understanding of complex extension libraries still needs to be improved.

In summary, this work provides an initial process design for LLM-assisted SysML v2 multi-source model collaboration and semantic alignment. Through a measurement system example, it demonstrates its operational feasibility and process stability, laying a replicable technical foundation for future research.

\bibliographystyle{IEEEtran}
\bibliography{reference.bib}

\end{document}